\documentclass[aps,floatfix,twocolumn,amsmath]{revtex4}
\usepackage{graphicx}
\newcommand{\be}{\begin{equation}}
\newcommand{\ee}{\end{equation}}
\newcommand{\bea}{\begin{eqnarray}}
\newcommand{\eea}{\end{eqnarray}}
\newcommand{\no}{\noindent}

\begin{document}


\title{Intersection Democracy for Winding Branes \\ and \\ Stabilization of Extra Dimensions}


\author{Tongu\c{c} Rador}
\email[]{tonguc.rador@boun.edu.tr}

\affiliation{Bo\~{g}azi\c{c}i University Department of Physics \\ 34342 Bebek, \.{I}stanbul, Turkey}


\date{\today}

\begin{abstract}

We show that, in the context of pure Einstein gravity, a democratic principle for intersection possibilities of branes winding around extra dimensions in a given partitioning yield stabilization, while what the observed space follows is matter-like dust evolution . Here democracy is used in the sense that, in a given decimation of extra dimensions, all possible wrappings and hence all possible intersections are allowed. Generally, the necessary and sufficient condition for this is that the  dimensionality $m$ of the observed space dimensions  obey $3\leq m \le N$ where $N$ is the decimation order of the extra dimensions. 

\end{abstract}


\maketitle

\section{Introduction}

One extension of standard theories of high energy physics possibly unifying all known forces  is string theory which mathematically necessitates extra dimensions other than the three we observe daily and that these extra dimensions are compact and very small. One important question once this proposition is accepted is that why these extra dimensions remained so small in contrast to the known universe. 
It is therefor important to look for ways to stabilize the size of extra dimensions within the context of cosmological evolution of the universe where it is known that the observed dimensions expand and have expanded throughout its history.

The literature on this nowadays rather vivid topic is considerable and we refer the reader to the following articles on brane gas cosmology \cite{art1}-\cite{art22}. Articles \cite{art1}-\cite{art6} particularly deal with the stabilization problem, while \cite{art7}-\cite{art22} are works on brane gas cosmology also relevant to this work. 

In this paper we confine the discussion to pure Einstein gravity and to brane winding modes only. 
The outline of the manuscript is as follows. After laying out the main mathematical formalism we present
two explicit cases of winding schemes. Namely the 2-fold and 3-fold decimations, where the extra dimensions
are divided into two and three lumps respectively. After discussing the possibilities for stabilization of extra dimensions in these explicit examples we then present stabilization conditions for an N-fold decimation where we made the simplifying assumption of symmetric decimation, that is the extra dimensions are divided
into N equidimensional parts, provided it can be done. This assumption makes it easy to deal with an otherwise very complicated non-linear coupled system. The general lesson is that for stabilization to occur
the number of dimensions of the observed space has to obey $3 \leq m \le N$. 
     
\section{Formalism}

The metric relevant for cosmological purposes is given by the following,
\be
ds^{2}=-dt^{2}+e^{2 B(t)} dx^{2} + \sum_{i} e^{2 C_{i}(t)} dy_{i}^{2}\;\;.
\ee
\no Here the $C_{i}$ and $y_{i}$ represent the scale factor and the coordinates of extra dimensions respectively, the dimensionality of each partition is $p_{i}$. For clarity we separated the observed dimensions with scale factor $B$ and dimensionality $m$. We also assume that the observed dimensions $x_{i}$ are non-compact  
and the $y_{i}$ represent compact extra dimensions with the topology of $S^{1}$ each. The total space time dimensionality is $d=1+m+\sum_{i}p_{i}$.

The pure Einstein gravity equations coupled to matter is
\be
R_{\mu\nu}-\frac{1}{2}R g_{\mu\nu}=\kappa^{2} T_{\mu\nu} \;.
\ee

\no With these assumptions the equations of motion for the scale factors can be cast as follows (we set $\kappa^{2}=1$),

\begin{subequations}
\bea
\dot{A}^2 &=& m \dot{B}^2 +\sum_{i} p_{i} \dot{C_{i}}^2 + 2\rho \;, \label{aninki}\\
\ddot{B}+\dot{A} \dot{B} &=& T_{\hat{b} \hat{b}}-\frac{1}{d-2} T \;,\\
\ddot{C_{i}}+\dot{A} \dot{C_{i}} &=& T_{\hat{c}_{i}\hat{c}_{i}} - \frac{1}{d-2} T\;, \\
A &\equiv& m B +\sum_{i} p_{i} C_{i} \;. 
\eea
\end{subequations}

\no The hatted indices refer the the orthonormal co-ordinates. Also $\rho$ represents the total energy density and $T_{\hat{\mu}\hat{\nu}}$ are the components of the total energy-momentum tensor while $T$ is its trace.

For brane winding modes the total energy momentum tensor for a particular winding pattern can be shown to be a sum of dust-like energy momentum tensors with pressure coefficients to be zero wherever there is no wrapping and minus one wherever there is wrapping \cite{art1}-\cite{art4}. 
The energy density for any such conserved energy-momentum tensor would be

\be
\rho^{\alpha}=\rho_{0}^{\alpha} \exp\left[-m B +\sum(1+\omega_{i} C_{i})\right]\;,
\ee

\no with $\rho_{0}^{\alpha}>0$ and as mentioned $w_{i}=-1$ for directions where there is a wrapped brane and $w_{i}=0$ if there is no brane wrapped in that direction. Following \cite{art4} we call $w_{i}=-1$ winding and $w_{i}=0$ transverse directions respectively. Note in particular that since branes only wrap around extra dimensions,  observed space is transverse to those and the corresponding pressure coefficient vanishes: we have $T_{\hat{b}\hat{b}}=0$.

Now if stabilization ever happens the rest of the equations should remain compatible. Stabilization would require all $\ddot{C_{i}}$ and $\dot{C_{i}}$ to vanish and we therefor get the following relation (upon observing that $\sum_{i}p_{i} \left[\rm RHS\;\;of\;\;C_{i}\;\;equations\right]=0$),

\be
\left(\frac{m-1}{d-2}\right)\;T\;=\;-\rho\;.
\ee

\no This results in the following equations for $B$

\begin{subequations}
\label{beqq}
\bea
\ddot{B}+m\dot{B}^{2}&=&\left(\frac{1}{m-1}\right)\;\rho\;,\\
m(m-1) \dot{B}^{2} &=& 2\;\rho\;.
\eea
\end{subequations}

\no Here $\rho=e^{-mB}\times \sigma$. The constant $\sigma$ depends on values of the stabilized scale $C_{i}(0)$ and energy density factors $\rho^{\alpha}_{0}$. The equations (\ref{beqq}) for $B$ are congruent only if $e^{B}\propto t^{2/m}$ which is the evolution of presureless dust. This is to be expected since branes do not exert any pressure along the observed dimensions. The only remaining condition is 

\be\label{final}
\sigma = 2 e^{m B(0)} \left(\frac{m-1}{m}\right)\;.
\ee

\no which can be satisfied without problem since we still have the choice of $B(0)$.

Thus if we can somehow find a way to stabilize the extra dimensions. This solutions will not spoil the rest of the equations and we would be safe. From now on we will focus on the $C_{i}$ equations to study stabilization.

\subsection{2-Fold Decimation}

In this section we would like to reproduce the results presented in \cite{art3,art4} to clarify the formalism. We divide the spacetime in the way $1+m+p+q$, that is the extra dimensions are divided in a 2-fold partitioning. We
also use the following winding scheme,

\[
(p)q\oplus p(q)
\]

The above is meant to read there is one brane wrapping along the $p$ directions and another one wrapping along the $q$ directions. The $C_{i}$ equations are therefore

\bea
-\frac{m+q-2}{d-2}\rho_{0}^{p} e^{-qC_{q}} + \frac{1+q}{d-2}\rho_{0}^{q} e^{-pC_{p}} &=& 0\;,\\
\frac{1+p}{d-2}\rho_{0}^{p} e^{-qC_{q}}-\frac{m+p-2}{d-2}\rho_{0}^{q} e^{-pC_{p}} &=& 0\;.
\eea

\no Which could be written as a matrix equation

\be\label{det1}
\left[\begin{tabular}[c]{cc}
$-(m+q-2)\;$ $1+q$ \\
\\
$1+p$ $-(m+p-2)$ 
\end{tabular}\right]\left[\begin{tabular}[c]{c} X\\ \\ Y\end{tabular}\right]=0\;.
\ee

\no Here we have defined $X=\rho_{0}^{p} e^{-qC_{q}}$ and $Y=\rho_{0}^{q} e^{-pC_{p}}$ and omitted the irrelevant factors. For a non-trivial solution we must have determinant of the matrix in (\ref{det1}) vanish. This quantity is -(m-3)(d-2). Therefor the necessary requirement is $m=3$. On the other hand however the solutions must all be positive definite as evident from the definitions of $X$ and $Y$. With $m=3$ the nullspace of the matrix in (\ref{det1}) is $(1,1)$. Therefor there is stabilization and necessary and sufficient condition is that $m=3$. 

We could enlarge the winding scheme to the following

\[
(p)q\oplus p(q)\oplus (pq)
\]

\no and this would give the following 

\be
\left[\begin{tabular}[c]{cc}
$-(m+q-2)\;$ $1+q$ \\
\\
$1+p\;$ $-(m+p-2)$ 
\end{tabular}\right]\left[\begin{tabular}[c]{c} X\\ \\ Y\end{tabular}\right]=(m-2) \rho^{pq}_{0}\left[\begin{tabular}[c]{c} 1\\ \\1\end{tabular}\right]
\ee

This would result in $X=Y=-(m-2)/(m-3)\rho_{0}^{pq}$ and again a positive definite solution would not be possible unless $m=3$. Therefor the winding mode $(pq)$ is forbidden for stabilization in this case.

\subsection{3-Fold Decimation}

To get further acquainted with the formalism let us consider a partitioning of the form $1+m+p+q+r$. And
consider the following cascaded winding scheme

\[
(pq)r\oplus p(qr) \oplus q(rp)
\]

\no With these assumptions the stabilization equations for extra dimensions read
\begin{widetext}
\be\label{det2}
\left[\begin{tabular}[c]{ccc}
$-(m+r-2)\;$ $1+q+r\;$ $-(m+q-2)$ \\
$-(m+r-2)\;$ $-(m+p-2)\;$ $1+p+r$\\
$1+p+q\;$ $-(m+p-2)\;$ $-(m+q-2)$
\end{tabular}\right]\left[\begin{tabular}[c]{c} X\\ Y\\Z\end{tabular}\right]= 0\;.
\ee

\no With $X=\rho_{0}^{pq}e^{-rC_{r}}$ and similarly for the others. The determinant of the matrix in (\ref{det2}) is $(5-2m)(d-2)^2$, which won't allow a non-trivial solution for integer $m$. To be able to go around this we could add the $(pqr)$ winding mode so that the total winding scheme is

\[
(pq)r\oplus p(qr) \oplus q(rp)\oplus (pqr)
\]

\no for which the stabilization equations are,

\be
\left[\begin{tabular}[c]{ccc}
$-(m+r-2)\;\;$ $1+q+r\;\;$ $-(m+q-2)$ \\
$-(m+r-2)\;\;$ $-(m+p-2)\;\;$ $1+p+r$\\
$1+p+q\;\;$ $-(m+p-2)\;\;$ $-(m+q-2)$
\end{tabular}\right]\;\left[\begin{tabular}[c]{c} X\\ Y\\Z\end{tabular}\right]= (m-2)\rho_{0}^{pqr}\left[\begin{tabular}[c]{c} 1\\1\\1\end{tabular}\right]\;.
\ee
\end{widetext}

\no The solutions to this system would be $X=Y=Z=\rho_{0}^{pqr} (m-2)/(5-2m)$ which makes it impossible to have positive definite solutions for $m\geq2$.

The only possibility left is to consider further winding modes around one part of the decimation. That is we
now look for the following winding scheme 

\[
(pq)r\oplus p(qr) \oplus q(rp)\oplus (pqr) \oplus (p)qr \oplus p(q)r \oplus pq(r)
\]

\no This scheme will bring quadratic terms involving $XY$, $XZ$ and $YZ$. Since for example the winding mode
$(p)qr$ will have two transverse partitions the corresponding energy density will be $\rho_{0}^{p} e^{-mB-qC_{q}-rC_{r}}$ which is proportional to $XY$ apart from the factor $e^{-mB}$ which will cancel out
in the stabilization equations. So this system will in principle be complicated. Nevertheless we can devise
a simpler way to procede as follows. The equations are invariant under the trivial permutations of quantities
depending on $p$, $q$ and $r$. Therefore if a stabilizing solutions exists generally, it should also exist for $p=q=r$ and with this we should take all the corresponding coefficients of energy densities to be equal since the only parameter these can depend are the dimensionality of the space around which they wrap (and topology which we took equivocally to be tori from the outset). For example the modes $(pq)r$, $p(qr)$ and $q(rp)$ will have equal strengths when $p=q=r$. This is what we would like to call
symmetric decimation and although it is not the general case it would give a hint on the general solution. For example if after setting $p=q=r$ the solution does not depend on $p$ we can argue as follows. Assume a non-symmetric solution exists and there is stabilization, then all the differences in the stabilization values of the scale factors of extra dimensions can be gauged away by rescaling the corresponding dimensions
by appropriate amounts. Thus a symmetric solution hints strongly to a solution in the general case. 

This argument simplifies matters to a considerable extent since we will only have one variable to deal with and the problem will resolve to the finding positive definite roots to a polynomial equation. For our winding scheme this equation will be

\be
\alpha (4-m)X^{2} + (5-2m) X +\beta (2-m) =0\;.
\ee

\no Here $\alpha$ and $\beta$ are positive numbers and are related to the energy densities of the corresponding winding modes. There are no positive definite solutions for the above equations for all positive $\alpha$ and $\beta$ unless $m=3$. This actually does not depend on the existence of $\beta$, provided the linear term in $X$ is present.

Thus we have shown that in 3-fold decimation a democratic winding scheme and hence a democratic intersection scheme is stabilizing the extra dimensions with positive real scale factors provided the dimensionality of the observed space is constrained as $m=3$.   

\subsection{N-Fold Symmetric Decimation}

The formalism in the previous part can be generalized easily to an N-fold democratic decimation and following the argument we presented we will consider symmetric decimation.

\bea
1+m+&\underbrace{p+p+p+.....}& \nonumber\\
&{\rm N\;\;factors}& \nonumber
\eea

\no the winding scheme will consist of 

\bea
(p)pppp... &\oplus& \;\; {\rm N sources}\oplus \nonumber\\
(pp)ppp... &\oplus& \;\; {\rm N sources}\oplus \nonumber\\
(ppp)pp... &\oplus& \;\; {\rm N sources}\oplus \nonumber\\
{\rm all \;\; N-1\;\;modes}&\vdots& \;\;{\rm N sources}\oplus \nonumber\\
 (ppp....) && \;\; {\rm 1 source}\nonumber
\eea

\no and this will lead to the following stabilization condition

\be\label{DAEQ}
P_{s}(X)\equiv\sum_{n=1}^{N-1} \alpha_{n} \xi_{n} X^{n}-\beta (m-2)=0\\
\ee

\no with 

\begin{subequations}
\label{DAEQ2}
\bea
\xi_{n} &=& (2N-n)-(N-n)m \\
\alpha_{1} &=& 1 \\
\alpha_{i} &\geq& 0; \\
\beta &\geq& 0;
\eea
\end{subequations}

Stabilization requires finding the solutions of (\ref{DAEQ}) for all $\alpha_{i}$ and $\beta$ such that $X>0$. For our purposes we do not need to know the most general solutions of polynomials of arbitrary order. We just need to find the condition on $m$ such that there are positive roots. To study this we just need to count the sign changes in the polynomial starting from the highest order term, then Descartes's sign rule tells us that the number of positive roots is either equal to the number of sign changes in the coefficients or less than it by a multiple of 2. The sign of the coefficients are determined by $\xi_{n}$ since all $\alpha_{i}\geq0$. One important thing to realize is that the coefficients $\xi_{n}$ are monotonically decreasing with $n$ for a given $m$. Thus, if there will be a sign change after all it will only occur once starting with the lowest degree term and the sign changing term changes and moves towards the highest degree term with increasing $m$. It can be shown that there are no sign changes for $m<3$ and $m\geq N$ (the upper bound start to operate for $N\geq 3$) and that there is only one sign change in between, meaning that there is only one positive root for all values of $\alpha_{n}$. Again this is independent of the full winding mode $(pppp....)$, provided we have the term linear in $X$, since the coefficient of the $\beta$ term is $(2-m)$ which always starts to change sign with the linear term. Unfortunately
this procedure does not fix $m$ unambiguously but it certainly puts a strict lower bound of $m\geq 3$ and a
decimation dependent upper bound of $m\leq N$ which is enough to state that $m=3$ is always sufficient to ensure stabilization although it is not necessary. One could turn the argument around and actually pick the decimation number to be $N=3$ which is the smallest number of decimations for which the bound operates and this unambiguously fixes $m$.  Finally the fact that there is only one positive root
hints at the possibility that the solution exists even for non-symmetric decimation.

To complete the argument we should remember that the equation (\ref{final}) be satisfied as well. With the 
parameters defined in (\ref{DAEQ}) and (\ref{DAEQ2}) this will read

\be
\beta + N\sum_{i}^{N-1} \alpha_{i}X^{i} = 2 e^{m B(0)} \left(\frac{m-1}{m}\right)\;.
\ee

\no which can be satisfied without problem by an appropriate choice of $B(0)$.

\subsection{Stability of the equilibrium point}

In this section we discuss that the stability point is in fact stable, that is all admissible (see below) initial data will converge to the stabilization point. We first show that the equilibirum 
point is linearly stable and then present an argument to show that stabilization is also achieved in the
non-linear regime.

\be\label{equic}
\ddot{C}=-\dot{A}\dot{C}+\frac{1}{d-2} e^{-mB} P_{s}(X=e^{-pC}).
\ee 
If we expand $C$ around the equilibrium point such that we keep only the linear perturbations $C=C_{0}+\delta{C}$ we would get the following

\be\label{equib}
\delta\ddot{C}=-\dot{A}\delta\dot{C}-\frac{p\delta C}{d-2}e^{-mB}\;X_{0}\; P_{s}'(X_{0}),.
\ee
where $P_{s}'(X_{0})$ denotes derivative of $P_{s}$ as the equilibrium point.

This is like a motion under two forces. The force which is proportional to $\dot{A}$ can either be a driving
or a friction force depending on the sign of $\dot{A}$ but as clear from the equations of motions ${\rm Sign}(\dot{A})$  is a constant of motion  since $\dot{A}$ is never allowed to vanish by  (\ref{aninki}). In \cite{art1} it has been observed that when ${\rm Sign}(\dot{A})<0$ one has a driving force and a singularity is reached in finite proper time. For ${\rm Sign}(\dot{A})>0$ on the other hand one has a friction-like force, this already is a hint for the stability of the equations in the long term. The other force is like a linear force
depending on the sign of the RHS of (\ref{equib}). This sign is unambiguously fixed by the requirements for
stabilization. Let us remember that the coefficients of $P_{s}(X)$ start becoming negative with increasing $m$ starting from the lowest order term (including the constant term), and we have shown that there is a unique positive root of $P_{s}(X)$ if there is at least one sign change in the coefficients. Thus the requirement for a unique non-degenerate positive root also requires that the sign of the coefficient of the highest order term be positive, this would mean that the derivative of the stabilization polynomial at the equilibirum point is positive since the positive root is the largest root and the polynomial will either increase or decrease depending on the sign of the coefficient of the largest order term . Therefor
the sign of the RHS of (\ref{equib}) is negative. The second force is an attractive linear force in the linear approximation. Hence the $C$ equations are linearly stable near the stabilization fixed point.

To try to understand what could happend in the non-linear regime let us remember again we have shown
,given the conditions on $m$, that there is a unique positive root $X_{0}$ of $P_{s}(X)$. This would mean that we  have $P_{s}(X)<0$ for $0<X<X_{0}$ and $P_{s}(X)>0$ for $X>X_{0}$ since the coefficient of the highest order term is positive. We will confine ourselves to   ${\rm Sign}(\dot{A})>0$. 

{\bf Case A:} Let us pick a point in the region $0<X<X_{0}$ as our initial data point. Here $\ddot{C}$ is initially negative if initially $\dot{C}>0$ ($C$ getting larger $X$ getting smaller) and hence $\dot{C}$ will become negative at some time in the future (since $\ddot{C}$ will never change sign until $\dot{C}$ becomes negative), when this transition happens we would have $C$ getting smaller and $X$ getting larger so this initial data eventually turns toward the equilibrium point $X_{0}$ with an initial data equivalent to, $0<X<X_{0}$ and $\dot{C}<0$ thus we should only consider this. In this case the sign of $\ddot{C}$ depends on the interplay between the terms in (\ref{equic}). There are two cases: either the evolution of the system
monotonically approaches $X_{0}$ with $\dot{C}$ approaching zero from below or passes this point with $\dot{C}<0$ and becomes a problem of Case B below.

{\bf Case B:} Let us pick a point in the region $X>X_{0}$ as our initial data point. Here $\ddot{C}$ is initially positive if initially $\dot{C}<0$ ($C$ getting smaller $X$ getting larger) and hence $\dot{C}$ will become
positive at some time in the future (since $\ddot{C}$ will never change sign until $\dot{C}$ becomes positive), when this transition happens we would get $C$ getting larger and $X$ getting smaller to this initial data eventually turns towars the equilibrium point $X_{0}$ with an initial data equivalent to
$X>X_{0}$ and $\dot{C}>0$ thus we should only consider this.  In this case the sign of $\ddot{C}$ depends on the interplay between the terms in (\ref{equic}). There are two cases: either the evolution of the system
monotonically approaches $X_{0}$ with $\dot{C}$ approaching zero from above or passes this point with $\dot{C}>0$ and becomes a problem of Case A above.

The considerations above are enough to argue that the point $X_{0}$ is an attractive fixed point: the solutions will converge to this point in the future.

\section{Conclusions}

In this work we have shown that in pure Einstein gravity the size of extra dimensions can be stabilized by winding modes alone in a particular winding system we called democratic winding scheme where intersections and windings of all possible kinds are allowed. This is an admirable result since during very early times in the cosmological evolution the extreme hot environment of the universe could somehow been incapable of choosing among winding schemes\footnote{the case of different topology is no concern in the present work since we assumed the same topology for all the internal dimensions from the outset.}. 

This model is incomplete in the sense that we have made the symmetric decimation choice and it is hard to infer directly what would happen for non-symmetric generalization. One possibly related  problem is about the Branderberger-Vafa (BV) \cite{art23} mechanism which implies that all $p$-branes with $p>2$ annihilated very early in the universe. To connect the BV construct to the model of this paper we first realize that there are only four symmetric decimations of the six extra dimensions of string theory. These are $111111$, $222$, $33$ and $6$. Only the first two are allowed in the BV formalism to have non-trivial wrapping and unfortunately allowing only branes with $p\leq 2$ yields bounds for $m$ strictly greater than three. Perhaps a variant of the model we presented can circumvent this.

There could be interesting extensions of the democratic winding idea with increasing
complexity. For example it is fairly easy to see that adding momentum excitations along winding directions would change things drastically since the equations will be generally polynomials with non-integer powers\footnote{this is expected because the pressure coefficient of momentum modes of a $p$-brane is $1/p$ in contrast to $p$ independent winding pressure which is $-1$.}. One could also try to enlarge the present system to dilaton gravity. Works along these lines are in progress.

\end{document}